\documentclass{article}
\usepackage{amsfonts}
\usepackage{amsmath}

\begin{document}

\title{A Variational Formulation of Classical Cosserat Elasticity with
Independent Coframe and Rotational Connection}
\author{Lev Steinberg \\
University of Puerto Rico at Mayaguez }
\maketitle

\begin{abstract}
We present a geometric formulation of classical Cosserat elasticity in which
the coframe and rotational connection are treated as independent variational
fields. In contrast to conventional metric-based approaches, this
formulation makes the underlying geometric structure explicit and separates
translational and rotational degrees of freedom at the level of the action.
The governing equations are obtained directly as Euler--Lagrange equations
and yield the Cosserat force and moment balance laws without imposing
compatibility constraints a priori.It is further shown that configurational
balances arise from invarianceof the action under material translations and
rotations via Noether's theorems, providing an explicit variational
interpretation of micropolar mechanics. A metric-free linearization recovers
the classical strain and wryness measures and establishes equivalence with
standard tensorial formulations under appropriate constitutive assumptions.
The proposed framework clarifies the role of the connection field, which
remains implicit in classical theories, and provides a geometrically
explicit variational framework.for Cosserat continua.The formulation also
provides a natural foundation for generalized incompatible Cosserat continua
and mesoscopic defect theories
\end{abstract}

\section{Introduction}

Cosserat continua provide one of the most important extensions of classical
elasticity by incorporating independent rotational degrees of freedom at
each material point. The classical theory introduced by E. and F. Cosserat 
\cite{Cosserat} assumes that every material point carries an orthonormal
triad describing microscopic orientation. This enrichment naturally leads to
asymmetric force stresses and the appearance of couple stresses in the
balance equations. Subsequent developments by Mindlin and Tiersten \cite%
{Mindlin} established a rigorous continuum formulation of micropolar
elasticity, while the work of Eringen \cite{Eringen} developed a unified
theory of microcontinua encompassing micropolar, microstretch, and
micromorphic models within a consistent constitutive framework. Despite
their success, most formulations of Cosserat elasticity remain tensorial and
metric-based. In these approaches, the geometric role of the connection
field is typically implicit, and the variational structure is not formulated
in terms of independent geometric variables. Differential geometry provides
a natural framework for describing continua with microstructure. The
geometric ideas introduced by Cartan \cite{Cartan}, Cartan \cite{Cartan 2},
based on coframes, connections, torsion, and curvature, offer a
coordinate-free description of continuum kinematics. These concepts were
later applied to defect mechanics by Kondo \cite{Kondo 1952}, Bilby et al. 
\cite{Bilby 1955}, and Kr\"{o}ner \cite{Kroner}, who interpreted torsion and
curvature as measures of dislocation and disclination densities.

These geometric ideas have recently been revisited in modern formulations of
continuum mechanics. Yavari and Goriely \cite{Yavari} developed a
Riemann--Cartan geometric framework for nonlinear dislocation mechanics,
while N\'{e}meth and Adhikari \cite{Nemeth} formulated Cosserat elasticity
using differential forms in a coordinate-free setting \ Although these
approaches highlight the effectiveness of geometric methods, they do not
adopt a Palatini-type variational framework \cite{Palatini}, in which the
coframe and connection are treated as independent fields.

The present work develops a geometric formulation of Cosserat elasticity
based on a Palatini variational principle which makes the underlying
geometric structure explicit. The governing equations are obtained as
Euler--Lagrange equations of the action and yield the Cosserat force and
moment balance laws without imposing compatibility conditions a priori.
Furthermore, Noether's theorem \cite{Noether} provides a direct
interpretation of these balances as consequences of invariance under spatial
translations and rotations.

The independent variational treatment of the connection employed here bears
structural similarities to formulations developed in relativistic field
theory, particularly in the work of Datta \cite{Datta 1971} and Hehl \cite%
{Helh 1976}. Although developed in a different physical context, these
formulations yield angular momentum balance structures analogous to
equilibrium conditions derived by Kr\"{o}ner \cite{Kroner 1968} and Stojanovi%
\'{c} \cite{Stojanovic 1969} for Cosserat continua. Analogous stress and
couple-stress structures were later employed by Edelen \cite{Edelen1988} in
geometric defect theory.

However, the present theory differs fundamentally in physical interpretation
and purpose. Here the coframe and connection describe the kinematics of a
generalized Cosserat continuum rather than spacetime gravitation, and the
resulting balance laws are interpreted as mechanical force- and
couple-stress balances within generalized defect mechanics. From a geometric
viewpoint, the coframe and rotational connection play geometric roles
associated with local translations and rotations, while torsion and
curvature represent the corresponding field strengths. In the present
context, however, these quantities describe the material fabric rather than
the geometry of physical space.

The present work should be viewed as the foundational variational level of a
broader hierarchy of generalized Cosserat defect theories. In the present
paper, the coframe and connection are treated as independent variables
within a Palatini-type variational framework, yielding the corresponding
force and moment balances. In subsequent extensions, torsion and curvature
may themselves be promoted to independent defect fields, leading naturally
to mesoscopic theories of distributed defects and to higher-order source
theories describing defect generation and annihilation.

The importance of the independent variational structure lies not merely in
geometric reformulation, but in separating translational and rotational
mechanics at the level of the action. This separation provides a natural
variational foundation for generalized incompatible Cosserat continua and
defect theories, see for example, [\cite{Steinberg2026}].

\section{Moving Manifolds and Motion}

Let $M$ be a smooth three-dimensional differentiable manifold representing
the material body in a reference description, and let $T\subset \mathbb{R}$
be a time interval of observation. The time-dependent body is described by
the space--time manifold 
\begin{equation*}
B=M\times T.
\end{equation*}%
Its points $(X,t)\in B$ label material points $X\in M$ together with time $%
t\in T$.

For each fixed $t$, the slice 
\begin{equation*}
M_{t}=M\times \{t\}
\end{equation*}%
represents the body at time $t$, so that the family $\{M_{t}\}_{t\in T}$ may
be interpreted as a moving manifold.

The physical motion of the body is described by a smooth embedding 
\begin{equation*}
\chi _{t}:M\rightarrow \mathbb{R}^{3},\qquad x=\chi (X,t),
\end{equation*}%
which assigns to each material point $X\in M$ its spatial position $x\in 
\mathbb{R}^{3}$ at time $t$. For each fixed $t$, the image 
\begin{equation*}
\chi _{t}(M)\subset \mathbb{R}^{3}
\end{equation*}%
is the current configuration.

The time derivative of the motion defines the material velocity field 
\begin{equation*}
v(X,t)=\frac{\partial \chi }{\partial t}(X,t),
\end{equation*}%
which induces the usual Eulerian velocity field on the current configuration.

The differential of the motion with respect to $X$ gives the deformation
gradient 
\begin{equation*}
F(X,t)=D_{X}\chi (X,t):T_{X}M\rightarrow T_{x}\mathbb{R}^{3}.
\end{equation*}%
In a frame--coframe description this maps a reference coframe $\{E^{A}\}$ on 
$M$ to a current coframe $\{e^{i}\}$ on the current configuration: 
\begin{equation*}
e^{i}(X,t)=F^{i}{}_{A}(X,t)\,E^{A}(X).
\end{equation*}

In classical non-Cosserat elasticity all geometric quantities are derived
from the motion $\chi$, and the underlying connection is determined by the
induced metric. Consequently torsion and curvature are constrained to
vanish, so that dislocations and disclinations cannot appear as independent
geometric fields.

\section{Geometric Structure of Cosserat Continuum}

Let $M_{t}$ be equipped with a coframe 
\begin{equation*}
e^{i}\in \Omega ^{1}(M_{t})
\end{equation*}%
and a rotational connection one-form 
\begin{equation*}
\omega ^{i}{}_{j}\in \Omega ^{1}(M_{t};\mathfrak{so}(3)).
\end{equation*}%
The associated torsion and curvature are 
\begin{equation*}
T^{i}=De^{i}=de^{i}+\omega ^{i}{}_{j}\wedge e^{j},
\end{equation*}%
\begin{equation*}
\Omega ^{i}{}_{j}=D\omega ^{i}{}_{j}=d\omega ^{i}{}_{j}+\omega
^{i}{}_{k}\wedge \omega ^{k}{}_{j}.
\end{equation*}

In the Palatini formulation the coframe and the connection are treated as
independent fields in the variational principle 
\begin{equation*}
\delta S[e,\omega ]=0,
\end{equation*}%
with independent variations $\delta e^{i}$ and $\delta \omega ^{i}{}_{j}$.
Compatibility relations such as vanishing torsion arise as field equations
or constitutive restrictions rather than being imposed a priori.

\section{Classical Cosserat Elasticity}

In the compatible defect-free classical case we write 
\begin{equation*}
e^{i}(X,t)=F^{i}{}_{A}(X,t)\,E^{A},
\end{equation*}%
and impose the torsion-free and curvature-free conditions 
\begin{equation*}
T^{i}=De^{i}=0,\qquad \Omega ^{i}{}_{j}=D\omega ^{i}{}_{j}=0.
\end{equation*}

If, in addition, one works in a local Cartesian frame with vanishing
connection, then 
\begin{equation*}
de^{i}=0.
\end{equation*}%
By the Poincar\'{e} lemma, locally there exist functions $y^{i}$ such that 
\begin{equation*}
e^{i}=dy^{i}=F^{i}{}_{A}\,dX^{A}.
\end{equation*}%
Hence 
\begin{equation*}
F^{i}{}_{A}=\partial _{A}y^{i},
\end{equation*}%
and the compatibility condition becomes 
\begin{equation*}
\partial _{[C}F^{i}{}_{A]}=0.
\end{equation*}%
Thus classical elasticity appears as the compatible special case of the
geometric framework. \ The main characteristic feature of Cosserat
elasticity is that the coframe and rotational structure are treated
independently. In the defect-free classical Cosserat regime one still has 
\begin{equation*}
T^{i}=0,\qquad \Omega ^{i}{}_{j}=0,
\end{equation*}%
but the rotational field is no longer determined solely by the deformation.
If the connection is flat ($\Omega ^{i}{}_{j}=0)$, it is locally pure gauge.
Thus one may write 
\begin{equation*}
\omega =Q^{-1}dQ,\qquad Q(X,t)\in SO(3).
\end{equation*}%
Let $\bar{e}^{k}$ denote a reference or elastic coframe,. then the current
coframe may be written as 
\begin{equation*}
e^{i}=Q^{i}{}_{k}\,\bar{e}^{k}.
\end{equation*}%
This representation makes clear that the rotational microstructure is
independent of the translational deformation, even though both enter the
classical elastic response.

\section{Lie Derivative of the Coframe and Infinitesimal Variations}

Assume an orthonormal coframe, 
\begin{equation*}
g=\delta _{ij}e^{i}\otimes e^{j}.
\end{equation*}%
Let $u=u^{j}\partial _{j}$ be a displacement vector field generating a flow $%
\Phi _{t}$. The infinitesimal variation of any tensor $H$ under this flow is 
\begin{equation*}
\delta H=\left. \frac{d}{dt}\Phi _{t}^{\ast }H\right\vert _{t=0}=L_{u}H.
\end{equation*}

Then applying Cartan's formula to the coframe gives 
\begin{equation}
L_{u}e^{i}=d(i_{u}e^{i})+i_{u}(de^{i}).  \label{f1}
\end{equation}%
Since 
\begin{equation*}
i_{u}e^{i}=u^{i},
\end{equation*}%
we obtain 
\begin{equation*}
d(i_{u}e^{i})=du^{i}.
\end{equation*}%
In the torsion-free case, 
\begin{equation*}
de^{i}=-\omega ^{i}{}_{j}\wedge e^{j}.
\end{equation*}%
Therefore 
\begin{align*}
i_{u}(de^{i})& =-\,i_{u}(\omega ^{i}{}_{j}\wedge e^{j}) \\
& =-(i_{u}\omega ^{i}{}_{j})e^{j}+\omega ^{i}{}_{j}(i_{u}e^{j}) \\
& =-(i_{u}\omega ^{i}{}_{j})e^{j}+\omega ^{i}{}_{j}u^{j}.
\end{align*}%
Hence substituting in \ref{f1} 
\begin{equation*}
L_{u}e^{i}=(du^{i}+\omega ^{i}{}_{j}u^{j})-(i_{u}\omega ^{i}{}_{j})e^{j}.
\end{equation*}%
Recognizing the covariant derivative 
\begin{equation*}
(Du)^{i}=du^{i}+\omega ^{i}{}_{j}u^{j},
\end{equation*}%
we obtain 
\begin{equation*}
L_{u}e^{i}=(Du)^{i}-\phi ^{i}{}_{j}e^{j},\qquad
\end{equation*}

where $\phi ^{i}{}_{j}$ denotes $i_{u}\omega ^{i}{}_{j}.$

Thus the Lie derivative decomposes into a translational part and an
infinitesimal rotational part. The local microrotation acts as 
\begin{equation*}
\delta _{\mathrm{rot}}e^{i}=\phi ^{i}{}_{j}e^{j},\qquad \phi _{ij}=-\phi
_{ji}.
\end{equation*}%
(((()(((In our three dimensional case , we can write 
\begin{equation*}
\phi ^{ij}=\varepsilon ^{ij}{}_{k}\,\varphi ^{k},
\end{equation*}%
and the infinitesimal coframe variation becomes 
\begin{equation*}
\delta e^{i}{}_{c}=\partial _{c}u^{i}-\varepsilon ^{i}{}_{jc}\varphi ^{j},
\end{equation*}%
which is the linearized Cosserat strain measure.

\section{Linearized Deformation}

Let $M$ be endowed with a reference coframe $E^{A}$. \ and define 
\begin{equation*}
e^{i}=F^{i}{}_{A}E^{A}
\end{equation*}%
be a time-dependent Cosserat coframe \ and $F^{i}{}_{A}$\ - deformation
gradient . Suppose that the deformation and microrotation are small, so that 
\begin{equation*}
y^{i}(X,t)=X^{i}+u^{i}(X,t),
\end{equation*}%
and 
\begin{equation*}
Q^{i}{}_{j}(X,t)\simeq \delta ^{i}{}_{j}-\varepsilon ^{i}{}_{jk}\varphi
^{k}(X,t).
\end{equation*}

Assume that 
\begin{equation*}
e^{i}=Q^{i}{}_{j}\,dy^{j}.
\end{equation*}

Then, to first order in $u^{i}{}_{,A}$ and $\varphi^{k}$, 
\begin{equation*}
F^{i}{}_{A} \simeq \delta^{i}{}_{A} + u^{i}{}_{,A} -
\varepsilon^{i}{}_{Ak}\varphi^{k}.
\end{equation*}

Equivalently, 
\begin{equation*}
F^{i}{}_{A} \simeq \delta^{i}{}_{A} + u^{i}{}_{,A} +
\varepsilon^{i}{}_{kA}\varphi^{k}.
\end{equation*}

Consequently, after identifying material and spatial indices in the
linearized setting, 
\begin{equation*}
\gamma_{ij} = F_{ij}-\delta_{ij} = u_{i,j}-\varepsilon_{ijk}\varphi_{k},
\end{equation*}
which is the classical micropolar strain tensor. Moreover, the corresponding
linearized wryness tensor is 
\begin{equation*}
\kappa_{ij}=\varphi_{i,j}.
\end{equation*}

\section*{Proof}

Since 
\begin{equation*}
y^{j}=X^{j}+u^{j},
\end{equation*}
we have 
\begin{equation*}
dy^{j}=dX^{j}+du^{j}.
\end{equation*}

With respect to the reference coframe $E^{A}$, 
\begin{equation*}
dX^{j}=\delta^{j}{}_{A}E^{A}, \qquad du^{j}=u^{j}{}_{,A}E^{A}.
\end{equation*}

Hence 
\begin{equation*}
dy^{j} = \left(\delta^{j}{}_{A}+u^{j}{}_{,A}\right)E^{A}.
\end{equation*}

Using 
\begin{equation*}
Q^{i}{}_{j} \simeq \delta^{i}{}_{j}-\varepsilon^{i}{}_{jk}\varphi^{k},
\end{equation*}
we obtain 
\begin{equation*}
e^{i} = Q^{i}{}_{j}dy^{j} \simeq \left(\delta^{i}{}_{j}
-\varepsilon^{i}{}_{jk}\varphi^{k}\right)
\left(\delta^{j}{}_{A}+u^{j}{}_{,A}\right)E^{A}.
\end{equation*}

Multiplying gives 
\begin{equation*}
e^{i} \simeq \left[ \delta^{i}{}_{j}\delta^{j}{}_{A} +
\delta^{i}{}_{j}u^{j}{}_{,A} -
\varepsilon^{i}{}_{jk}\varphi^{k}\delta^{j}{}_{A} -
\varepsilon^{i}{}_{jk}\varphi^{k}u^{j}{}_{,A} \right]E^{A}.
\end{equation*}

The last term, 
\begin{equation*}
-\varepsilon^{i}{}_{jk}\varphi^{k}u^{j}{}_{,A},
\end{equation*}
is second order, since it contains the product of the small microrotation $%
\varphi^{k}$ and the small displacement gradient $u^{j}{}_{,A}$. Therefore
it is neglected in the linear approximation. Thus 
\begin{equation*}
e^{i} \simeq \left( \delta^{i}{}_{A} + u^{i}{}_{,A} -
\varepsilon^{i}{}_{Ak}\varphi^{k} \right)E^{A}.
\end{equation*}

Since 
\begin{equation*}
e^{i}=F^{i}{}_{A}E^{A},
\end{equation*}
we identify 
\begin{equation*}
F^{i}{}_{A} \simeq \delta^{i}{}_{A} + u^{i}{}_{,A} -
\varepsilon^{i}{}_{Ak}\varphi^{k}.
\end{equation*}

Using antisymmetry, 
\begin{equation*}
-\varepsilon^{i}{}_{Ak}\varphi^{k} = \varepsilon^{i}{}_{kA}\varphi^{k}.
\end{equation*}

Therefore, 
\begin{equation*}
F^{i}{}_{A} \simeq \delta^{i}{}_{A} + u^{i}{}_{,A} +
\varepsilon^{i}{}_{kA}\varphi^{k}.
\end{equation*}

Lowering indices and identifying $A$ with $j$ gives 
\begin{equation*}
F_{ij}-\delta_{ij} = u_{i,j} - \varepsilon_{ijk}\varphi_{k}.
\end{equation*}

Hence 
\begin{equation*}
\gamma_{ij} = u_{i,j} - \varepsilon_{ijk}\varphi_{k}.
\end{equation*}

Finally, the first-order microrotation gradient defines the wryness tensor, 
\begin{equation*}
\kappa_{ij}=\varphi_{i,j}.
\end{equation*}

This completes the proof..

\section{Variational Formulation of Cosserat Elasticity}

We take spacetime to be the product manifold $M\times \mathbb{R}$. 

The action is 
\begin{equation*}
S[e,\omega]=\int_{\mathbb{R}}\int_{M}L(e^{i},\partial_{t}e^{i},%
\omega^{i}{}_{j},\partial_{t}\omega^{i}{}_{j}),
\end{equation*}
where $L$ is a Lagrangian density.

Define the conjugate stress and momentum forms 
\begin{equation*}
\Sigma _{i}:=\frac{\partial L}{\partial e^{i}}\in \Omega ^{2}(M),\qquad
M^{i}{}_{j}:=\frac{\partial L}{\partial \omega ^{j}{}_{i}}\in \Omega ^{2}(M),
\end{equation*}%
\begin{equation*}
P_{i}:=\frac{\partial L}{\partial (\partial _{t}e^{i})}\in \Omega
^{2}(M),\qquad Q^{i}{}_{j}:=\frac{\partial L}{\partial (\partial _{t}\omega
^{j}{}_{i})}\in \Omega ^{2}(M).
\end{equation*}

\subsection{Euler-Lagrange field equations}

Let the action be varied with respect to the independent fields $e^{i}$ and $%
\omega ^{i}{}_{j}$. Then the stationarity condition $\delta S=0$ implies the
field equations 
\begin{equation}
D\Sigma _{i}+\partial _{t}P_{i}=0,  \label{eq:forcebalance}
\end{equation}%
\begin{equation}
DM^{i}{}_{j}+e^{i}\wedge \Sigma _{j}+\partial _{t}Q^{i}{}_{j}=0.
\label{eq:momentbalance}
\end{equation}

These equations represent the force and moment balance laws of defect-free
Cosserat elasticity.

\paragraph{Derivation}

\subparagraph{Translational variation}

Let

\begin{equation}
\xi=\xi ^{i}e_{i}\in\Omega ^{0}(B_{0};\mathbb{R}^{3})
\end{equation}

be an arbitrary translational virtual field. The induced compatible
variations are

\begin{equation}
\delta _{\xi}e^{i}=D\xi ^{i}, \qquad \delta _{\xi}\omega ^{i}{}_{j}=0.
\label{translational-geometric-variations}
\end{equation}

The corresponding velocity variation is

\begin{equation}
\delta _{\xi}v^{i}=\partial _{t}\xi ^{i}.
\label{translational-velocity-variation}
\end{equation}

The internal translational term is

\begin{equation}
\int_{B_{0}}\Sigma _{i}\wedge D\xi ^{i}.
\end{equation}

Since $\Sigma _{i}$ is a two-form,

\begin{equation}
D(\Sigma _{i}\xi ^{i}) =(D\Sigma _{i})\xi ^{i} +\Sigma _{i}\wedge D\xi ^{i}.
\end{equation}

Therefore,

\begin{equation}
\int_{B_{0}}\Sigma _{i}\wedge D\xi ^{i} =\int_{\partial B_{0}}\Sigma _{i}\xi
^{i} -\int_{B_{0}}(D\Sigma _{i})\xi ^{i}.  \label{translational-space-parts}
\end{equation}

The kinetic term satisfies

\begin{equation}
P_{i}\partial _{t}\xi ^{i} =\partial _{t}(P_{i}\xi ^{i}) -(\partial
_{t}P_{i})\xi ^{i}.  \label{translational-time-parts}
\end{equation}

Assume that $\xi ^{i}$ vanishes at the initial and final times. After the
boundary traction is treated separately, the d'Alembert virtual-power
statement gives

\begin{equation}
\int_{B_{0}} (D\Sigma _{i}+\partial _{t}P_{i})\xi ^{i}=0.
\label{translational-virtual-power}
\end{equation}

The arbitrariness of $\xi ^{i}$ yields the dynamic force balance

\begin{equation}
D\Sigma _{i}+\partial _{t}P_{i}=0.  \label{dynamic-force-balance}
\end{equation}

Every term in (\ref{dynamic-force-balance}) is a three-form.

\subparagraph{Rotational variation}

Let

\begin{equation}
\lambda ^{i}{}_{j}=-\lambda ^{j}{}_{i}, \qquad \lambda\in\Omega ^{0}(B_{0};%
\mathrm{so}(3))
\end{equation}

be an arbitrary infinitesimal local rotation. The induced variations are

\begin{equation}
\delta _{\lambda}e^{i} =\lambda ^{i}{}_{j}e^{j},
\label{rotational-coframe-variation}
\end{equation}

and

\begin{equation}
\delta _{\lambda}\omega ^{i}{}_{j} =-D\lambda ^{i}{}_{j}.
\label{rotational-connection-variation}
\end{equation}

The associated angular-velocity variation is

\begin{equation}
\delta _{\lambda}\nu ^{i}{}_{j} =\partial _{t}\lambda ^{i}{}_{j}.
\label{rotational-velocity-variation}
\end{equation}

Both geometric variations are generated by the same local rotation. This
produces the moment of the force stress and the divergence of the couple
stress in one equation.

\subparagraph{Force-stress moment}

The coframe contribution is

\begin{equation}
\Sigma _{i}\wedge\delta _{\lambda}e^{i} =\lambda ^{i}{}_{j}e^{j}\wedge\Sigma
_{i}.
\end{equation}

Since $\lambda ^{i}{}_{j}$ is skew-symmetric, only the antisymmetric part
contributes:

\begin{equation}
\lambda ^{i}{}_{j}e^{j}\wedge\Sigma _{i} =\frac{1}{2}\lambda ^{i}{}_{j}
(e^{j}\wedge\Sigma ^{i}-e^{i}\wedge\Sigma ^{j}).
\label{antisymmetric-force-stress-term}
\end{equation}

With the mixed-index convention, the corresponding local force-stress moment
is represented by

\begin{equation}
e_{i}\wedge\Sigma ^{j} -e^{j}\wedge\Sigma _{i}.  \label{force-stress-moment}
\end{equation}

\subparagraph{Couple-stress divergence}

The connection contribution is

\begin{equation}
-\frac{1}{2}\int_{B_{0}} M_{i}{}^{j}\wedge D\lambda ^{i}{}_{j}.
\end{equation}

Since $M_{i}{}^{j}$ is a two-form,

\begin{equation}
D(M_{i}{}^{j}\lambda ^{i}{}_{j}) =(DM_{i}{}^{j})\lambda ^{i}{}_{j}
+M_{i}{}^{j}\wedge D\lambda ^{i}{}_{j}.
\end{equation}

Consequently,

\begin{equation}
-\frac{1}{2}\int_{B_{0}}M_{i}{}^{j}\wedge D\lambda ^{i}{}_{j} =\frac{1}{2}%
\int_{B_{0}}(DM_{i}{}^{j})\lambda ^{i}{}_{j} -\frac{1}{2}\int_{\partial
B_{0}}M_{i}{}^{j}\lambda ^{i}{}_{j}.  \label{couple-stress-integration}
\end{equation}

After the boundary couple traction is treated separately, the local
contribution is $DM_{i}{}^{j}$.

\subparagraph{Angular inertia}

The angular kinetic term satisfies

\begin{equation}
\frac{1}{2}Q_{i}{}^{j}\partial _{t}\lambda ^{i}{}_{j} =\frac{1}{2}\partial
_{t}(Q_{i}{}^{j}\lambda ^{i}{}_{j}) -\frac{1}{2}(\partial
_{t}Q_{i}{}^{j})\lambda ^{i}{}_{j}.  \label{angular-time-parts}
\end{equation}

Assume that $\lambda ^{i}{}_{j}$ vanishes at the initial and final times.
After discarding the temporal endpoint term and using the adopted
d'Alembert sign convention, the inertial contribution to the balance is

\begin{equation}
+\frac{1}{2}(\partial _{t}Q_{i}{}^{j})\lambda ^{i}{}_{j}.
\end{equation}

\paragraph{Dynamic moment balance}

After the spatial and temporal boundary terms are treated separately, the
rotational virtual-power statement is

\begin{equation}
\frac{1}{2}\int_{B_{0}} (DM_{i}{}^{j}+e_{i}\wedge\Sigma ^{j}
-e^{j}\wedge\Sigma _{i} +\partial _{t}Q_{i}{}^{j})\lambda ^{i}{}_{j}=0.
\label{rotational-virtual-power}
\end{equation}

The arbitrariness of the skew-symmetric field $\lambda ^{i}{}_{j}$ yields
the dynamic moment balance

\begin{equation}
DM_{i}{}^{j}+e_{i}\wedge\Sigma ^{j} -e^{j}\wedge\Sigma _{i}
+\partial _{t}Q_{i}{}^{j}=0.  \label{dynamic-moment-balance}
\end{equation}

Every term in (\ref{dynamic-moment-balance}) is a three-form.

Because $M^{i}{}_{j}$, $Q^{i}{}_{j}$, and $\lambda ^{i}{}_{j}$ are
$\mathfrak{so}(3)$-valued, the moment equation is understood after projection
onto its skew part.  It may therefore be written compactly as

\begin{equation}
DM^{i}{}_{j}+e^{i}\wedge\Sigma _{j}
+\partial _{t}Q^{i}{}_{j}=0. \label{compact-moment-balance}
\end{equation}

Equivalently, with both indices raised and antisymmetrization understood,

\begin{equation}
DM^{ij}+e^{[i}\wedge\Sigma ^{j]}
+\partial _{t}Q^{ij}=0. \label{skew-momentbalance}
\end{equation}

Thus (\ref{compact-moment-balance}) and
(\ref{skew-momentbalance}) are the same $\mathfrak{so}(3)$-valued balance;
the second formula merely displays the skew projection explicitly.

\subsubsection{Final dynamic system}

The physical dynamic force and moment balances are

\begin{equation}
D\Sigma _{i}+\partial _{t}P_{i}=0,
\end{equation}

and

\begin{equation}
DM_{i}{}^{j}+e_{i}\wedge\Sigma ^{j} -e^{j}\wedge\Sigma _{i}
+\partial _{t}Q_{i}{}^{j}=0.
\end{equation}

\section{Classical Micropolar Tensorial Form}

In three dimensions the balance equations written in differential-form
notation reduce to the standard tensorial equations of micropolar elasticity.

\subsection*{Force balance}

Write the force stress 2-form as 
\begin{equation*}
\Sigma _{i}=\frac{1}{2}(\Sigma _{i})_{ab}\,dx^{a}\wedge dx^{b},
\end{equation*}%
and define the dual force stress tensor by 
\begin{equation*}
(\Sigma _{i})_{ab}=\varepsilon _{abc}\sigma ^{c}{}_{i}.
\end{equation*}%
Let 
\begin{equation*}
P_{i}=p_{i}\,dV,\qquad F_{i}=f_{i}\,dV.
\end{equation*}%
Then the balance equation 
\begin{equation*}
D\Sigma _{i}=\partial _{t}P_{i}+F_{i}
\end{equation*}%
reduces in Euclidean coordinates to 
\begin{equation*}
\partial _{a}\sigma ^{a}{}_{i}=\partial _{t}p_{i}+f_{i}.
\end{equation*}%
If 
\begin{equation*}
p_{i}=\rho \,\dot{u}_{i},
\end{equation*}%
one obtains the classical linear momentum balance 
\begin{equation}
\partial _{a}\sigma ^{a}{}_{i}+f_{i}=\rho \,\ddot{u}_{i}.
\end{equation}

\subsection*{Angular momentum balance}

Write the couple-stress 2-form as 
\begin{equation*}
M^{i}{}_{j}=\frac{1}{2}(M^{i}{}_{j})_{ab}\,dx^{a}\wedge dx^{b},
\end{equation*}%
with 
\begin{equation*}
(M^{i}{}_{j})_{ab}=\varepsilon _{abc}\mu ^{c}{}_{ij},\qquad \mu
^{c}{}_{ij}=-\mu ^{c}{}_{ji}.
\end{equation*}%
Let 
\begin{equation*}
Q_{ij}=q_{ij}\,dV,\qquad C_{ij}=c_{ij}\,dV.
\end{equation*}%
Then the moment balance becomes 
\begin{equation*}
\partial _{a}\mu ^{a}{}_{ij}+\sigma _{ij}-\sigma _{ji}+c_{ij}=\partial
_{t}q_{ij}.
\end{equation*}

Since the pair $(ij)$ is skew-symmetric, it may be dualized in three
dimensions. Define 
\begin{equation*}
\mu ^{a}{}_{r}=\frac{1}{2}\varepsilon _{r}{}^{ij}\mu ^{a}{}_{ij},\qquad
\varphi _{r}=\frac{1}{2}\varepsilon _{r}{}^{ij}\phi _{ij},\qquad c_{r}=\frac{%
1}{2}\varepsilon _{r}{}^{ij}c_{ij}.
\end{equation*}%
Then the angular momentum balance takes the standard micropolar form 
\begin{equation}
\partial _{a}\mu ^{a}{}_{r}+\varepsilon _{r}{}^{ij}\sigma
_{ij}+c_{r}=J_{r}^{s}\,\ddot{\varphi}_{s}.
\end{equation}%
For isotropic microinertia $J_{r}^{s}=J\delta _{r}^{s}$ this reduces to%
\begin{equation*}
\partial _{a}\mu ^{a}{}_{r}+\varepsilon _{r}{}^{ij}\sigma _{ij}+c_{r}=J\,%
\ddot{\varphi}_{r}.
\end{equation*}

\section*{Configurational Balances}

The lassical Cosserat elasticity is treated as a compatible continuum
theory. The independent mechanical variables are the coframe $e^{i}$ and the
microrotation connection $\omega ^{i}{}_{j}$. The purpose of this theorem is
to show that, in addition to the usual physical balances of force and
moment, the same variational structure also gives two dynamic
configurational balances.

These balances are obtained from invariance of the action under material
translations and material rotations. They are not additional physical force
balances. Rather, they describe the balance of configurational force and
configurational moment associated with material relabeling and material
rotational invariance.below we formulate desribe the balances followed by
detailed proof.

Let $M$ be a material manifold endowed with a reference coframe $E^{A}$ and
the dual material frame $E_{A}$. Let the Lagrangian 3-form be 
\begin{equation*}
L=L(e^{i},\omega ^{i}{}_{j},\dot{e}^{i},\dot{\omega}^{i}{}_{j};X,t).
\end{equation*}

Define the configurational stress 2-form by 
\begin{equation*}
\mathcal{E}_A = i_{E_A}L - (i_{E_A}e^i)\Sigma_i -
(i_{E_A}\omega^i{}_{j})M^j{}_{i}.
\end{equation*}

and the configurational momentum 3-form by 
\begin{equation*}
\mathcal{P}_{A}=-(i_{E_{A}}e^{i})P_{i}-(i_{E_{A}}\omega
^{i}{}_{j})Q^{j}{}_{i}.
\end{equation*}

If the action is invariant under material translations, then the dynamic
configurational force balance is 
\begin{equation*}
D\mathcal{E}_A+\partial_t\mathcal{P}_A+\mathcal{F}_A=0,
\end{equation*}
where the configurational force 3-form is 
\begin{equation*}
\mathcal{F}_A=-i_{E_A}D_XL.
\end{equation*}

Here $D_XL$ denotes the explicit material dependence of the Lagrangian.
Therefore, if the material is homogeneous, then 
\begin{equation*}
D_XL=0,
\end{equation*}
and consequently 
\begin{equation*}
\mathcal{F}_A=0.
\end{equation*}

Thus, in the homogeneous Level 1 case, 
\begin{equation*}
D\mathcal{E}_A+\partial_t\mathcal{P}_A=0.
\end{equation*}

Here we define the configurational moment stress 2-form by 
\begin{equation*}
\mathcal{M}_{AB}=X_{A}\mathcal{E}_{B}-X_{B}\mathcal{E}_{A}+\mathcal{S}_{AB},
\end{equation*}%
where $\mathcal{S}_{AB}$ is the intrinsic configurational spin 2-form
associated with the Cosserat microstructure.

Define the configurational angular momentum 3-form by 
\begin{equation*}
\mathcal{Q}_{AB} = X_A\mathcal{P}_B - X_B\mathcal{P}_A + \mathcal{R}_{AB},
\end{equation*}
where $\mathcal{R}_{AB}$ is the intrinsic configurational angular momentum
3-form.

If the action is invariant under material rotations, then the dynamic
configurational moment balance is 
\begin{equation*}
D\mathcal{M}_{AB} + E_A\wedge \mathcal{E}_B - E_B\wedge \mathcal{E}_A +
\partial_t\mathcal{Q}_{AB} + \mathcal{C}_{AB} =0.
\end{equation*}

Here $\mathcal{C}_{AB}$ is the configurational moment source 3-form. It
measures the explicit lack of material rotational invariance. If the
material is homogeneous and isotropic in the material sense, then 
\begin{equation*}
\mathcal{C}_{AB}=0.
\end{equation*}

Therefore, in the homogeneous isotropic Level 1 case, 
\begin{equation*}
D\mathcal{M}_{AB} + E_A\wedge \mathcal{E}_B - E_B\wedge \mathcal{E}_A +
\partial_t\mathcal{Q}_{AB} =0.
\end{equation*}

Thus Level 1 classical Cosserat elasticity possesses two dynamic
configurational balances: 
\begin{equation*}
D\mathcal{E}_A+\partial_t\mathcal{P}_A=0,
\end{equation*}
and 
\begin{equation*}
D\mathcal{M}_{AB} + E_A\wedge \mathcal{E}_B - E_B\wedge \mathcal{E}_A +
\partial_t\mathcal{Q}_{AB} =0,
\end{equation*}
provided the material is homogeneous and isotropic in the material sense.

\section*{Proof}

The proof is based on Noether invariance of the action. Let 
\begin{equation*}
\mathcal{A}=\int L
\end{equation*}
be the action functional. The first variation of the Lagrangian has the form 
\begin{equation*}
\delta L = \delta e^i\wedge \Sigma_i + \delta\omega^i{}_{j}\wedge M^j{}_{i}
+ \delta\dot e^i\wedge P_i + \delta\dot\omega^i{}_{j}\wedge Q^j{}_{i} +
D(\cdots).
\end{equation*}

The dots denote boundary terms that arise from integration by parts. Here
the stress forms $\Sigma _{i}$ and $M^{i}{}_{j}$ are conjugates to $e^{i}$
and $\omega ^{i}{}_{j}$, while the momentum forms $P_{i}$ and $Q^{i}{}_{j}$
- conjugates to their time derivatives.

First consider an\textbf{\ infinitesimal material translation} generated by
the vector field 
\begin{equation*}
\xi =\xi ^{A}E_{A}.
\end{equation*}%
The induced variations of the fields are their material Lie derivatives, 
\begin{equation*}
\delta _{\xi }e^{i}=\mathcal{L}_{\xi }e^{i},\qquad \delta _{\xi }\omega
^{i}{}_{j}=\mathcal{L}_{\xi }\omega ^{i}{}_{j}.
\end{equation*}

The variation of the Lagrangian under a material translation can be written
in Noether form as 
\begin{equation*}
\delta_{\xi}L = D\left(\xi^A\mathcal{E}_A\right) + \partial_t\left(\xi^A%
\mathcal{P}_A\right) + \xi^A\mathcal{F}_A.
\end{equation*}

The spatial boundary current in this identity defines the configurational
stress: 
\begin{equation*}
\mathcal{E}_A = i_{E_A}L - (i_{E_A}e^i)\Sigma_i -
(i_{E_A}\omega^i{}_{j})M^j{}_{i}.
\end{equation*}

The time boundary current defines the configurational momentum: 
\begin{equation*}
\mathcal{P}_A = - (i_{E_A}e^i)P_i - (i_{E_A}\omega^i{}_{j})Q^j{}_{i}.
\end{equation*}

The remaining non-divergence contribution is caused by explicit material
dependence of the Lagrangian. Therefore one writes 
\begin{equation*}
\mathcal{F}_A=-i_{E_A}D_XL.
\end{equation*}

If the action is invariant under arbitrary material translations, then 
\begin{equation*}
\delta_{\xi}\mathcal{A}=0
\end{equation*}
for arbitrary $\xi^A$. Since $\xi^A$ is arbitrary, the coefficient of $\xi^A$
must vanish. This gives 
\begin{equation*}
D\mathcal{E}_A+\partial_t\mathcal{P}_A+\mathcal{F}_A=0.
\end{equation*}

This is the dynamic configurational force balance.

If the material has no explicit material inhomogeneity, then 
\begin{equation*}
D_XL=0.
\end{equation*}
Consequently, 
\begin{equation*}
\mathcal{F}_A=0,
\end{equation*}
and the balance becomes 
\begin{equation*}
D\mathcal{E}_A+\partial_t\mathcal{P}_A=0.
\end{equation*}

Now consider an \textbf{infinitesimal material rotation}. Let 
\begin{equation*}
\theta ^{AB}=-\theta ^{BA}
\end{equation*}%
be the antisymmetric material rotation parameter. The material coframe
varies as 
\begin{equation*}
\delta _{\theta }E^{A}=\theta ^{A}{}_{B}E^{B}.
\end{equation*}

The Noether current associated with this material rotational invariance is
the configurational moment current. Its spatial part is the configurational
moment stress 
\begin{equation*}
\mathcal{M}_{AB} = X_A\mathcal{E}_B - X_B\mathcal{E}_A + \mathcal{S}_{AB}.
\end{equation*}

The first two terms represent the orbital configurational moment generated
by the configurational stress. The term $\mathcal{S}_{AB}$ represents the
intrinsic configurational spin associated with the Cosserat microstructure.

The time component of the same Noether current is the configurational
angular momentum 
\begin{equation*}
\mathcal{Q}_{AB} = X_A\mathcal{P}_B - X_B\mathcal{P}_A + \mathcal{R}_{AB}.
\end{equation*}

The first two terms represent the orbital configurational angular momentum
generated by the configurational momentum. The term $\mathcal{R}_{AB}$
represents the intrinsic configurational angular momentum associated with
the Cosserat microstructure.

The variation of the Lagrangian under a material rotation can be written in
the Noether form 
\begin{equation*}
\delta_{\theta}L = D\left(\theta^{AB}\mathcal{M}_{AB}\right) +
\partial_t\left(\theta^{AB}\mathcal{Q}_{AB}\right) + \theta^{AB} \left(
E_A\wedge\mathcal{E}_B - E_B\wedge\mathcal{E}_A + \mathcal{C}_{AB} \right).
\end{equation*}

The term 
\begin{equation*}
E_A\wedge\mathcal{E}_B - E_B\wedge\mathcal{E}_A
\end{equation*}
is the orbital configurational moment generated by the configurational
stress. The term $\mathcal{C}_{AB}$ is the explicit material rotational
source. It vanishes if the material response is invariant under material
rotations.

Invariance of the action under arbitrary material rotations implies 
\begin{equation*}
\delta_{\theta}\mathcal{A}=0
\end{equation*}
for arbitrary antisymmetric $\theta^{AB}$. Therefore the coefficient of $%
\theta^{AB}$ must vanish, giving 
\begin{equation*}
D\mathcal{M}_{AB} + E_A\wedge \mathcal{E}_B - E_B\wedge \mathcal{E}_A +
\partial_t\mathcal{Q}_{AB} + \mathcal{C}_{AB} =0.
\end{equation*}

This is the dynamic configurational moment balance.

If the material is homogeneous and isotropic in the material sense, then 
\begin{equation*}
\mathcal{F}_A=0, \qquad \mathcal{C}_{AB}=0.
\end{equation*}
The two configurational balances reduce to 
\begin{equation*}
D\mathcal{E}_A+\partial_t\mathcal{P}_A=0,
\end{equation*}
and 
\begin{equation*}
D\mathcal{M}_{AB} + E_A\wedge \mathcal{E}_B - E_B\wedge \mathcal{E}_A +
\partial_t\mathcal{Q}_{AB} =0.
\end{equation*}

Thus material translation invariance gives the dynamic configurational force
balance, and material rotation invariance gives the dynamic configurational
moment balance. This completes the proof.

\section{Discussion}

The Palatini--Cosserat formulation developed in this work provides a
geometric and variational reinterpretation of classical Cosserat elasticity.
By treating the coframe and rotational connection as independent fields, the
formulation separates translational and rotational structures at the level
of the action and makes the underlying geometric framework explicit. In
contrast to conventional approaches, where compatibility conditions are
imposed a priori, these conditions arise naturally within the present
setting as constraints associated with the defect-free regime.

The variational structure leads directly to the Cosserat force and moment
balance laws as Euler--Lagrange equations and reveals their origin through
Noether's theorem as consequences of spatial invariance. This provides a
unified interpretation of micropolar mechanics in which balance laws are not
postulated but derived from symmetry principles.

From a geometric perspective, the formulation shares structural similarities
with gauge-theoretic formulations in which the coframe and connection act as
geometric variables associated with translations and rotations. However, in
the present context these fields describe the material fabric rather than
the geometry of physical space. This distinction is essential for the
interpretation of the theory as a continuum model of microstructure rather
than a geometric theory of spacetime.

The framework developed here also clarifies the transition from classical
Cosserat elasticity to mesoscopic theories of defects. In the defect-free
regime considered in this work, the conditions 
\begin{equation*}
T^{i}=0,\qquad \Omega ^{i}{}_{j}=0
\end{equation*}%
characterize compatible configurations. Relaxing these conditions leads
naturally to mesoscopic models in which torsion and curvature represent
dislocation and disclination densities. Furthermore, the same variational
structure admits identities associated with local symmetries, which in the
presence of defects give rise to configurational balance laws. These
extensions will be developed in subsequent work.

\section*{Conclusions}

A geometric formulation of classical Cosserat elasticity has been developed
within a Palatini variational framework in which the coframe and rotational
connection are treated as independent fields. The resulting Euler--Lagrange
equations yield the force and moment balance laws in a differential-form
representation, and Noether's theorem provides their interpretation as
consequences of spatial and material invariances of the action. A
metric-free linearization recovers the classical strain and wryness measures
and establishes consistency with the standard tensorial formulation of
micropolar elasticity. The present framework therefore provides a coherent
geometric and variational foundation for classical Cosserat elasticity. In
addition, the formulation establishes a natural point of departure for more
general theories in which torsion and curvature evolve as independent
defect-density fields These developments will be presented in subsequent
work.[\cite{Steinberg2026}]

The significance of the Palatini formulation does not lie merely in
rewriting classical Cosserat elasticity in differential-form language. Its
importance lies in establishing a variational framework in which
translational and rotational geometric structures are fundamentally
independent. The independent variational structure may be advantageous for
mixed numerical formulations involving translational and rotational degrees
of freedom. Although numerical implementations are beyond the scope of the
present work, the formulation appears compatible with finite-element
approaches designed for micropolar continua. \ The present formulation
therefore provides a geometrically explicit and variationally consistent
foundation for generalized Cosserat continua and their extension to
incompatible defect theories.

\end{document}